\documentclass[sigconf]{acmart}

\usepackage{forest}
\usepackage{tikz}
\usetikzlibrary{trees, matrix, patterns, positioning, arrows.meta}
\usepackage[utf8]{inputenc}
\usepackage{geometry}

\usepackage{multirow}
\usepackage{colortbl}
\usepackage{xcolor}
\usepackage{threeparttable}

\newtheorem{myDef1}{Definition}

\definecolor{connect-line}{RGB}{0,0,0}
\definecolor{middle-color}{RGB}{255,255,255}
\definecolor{leaf-color}{RGB}{255,255,255}
\definecolor{line-color}{RGB}{25,25,112}
\definecolor{extraction}{RGB}{240, 128, 128}
\definecolor{inversion}{RGB}{124, 205, 124}
\definecolor{others}{RGB}{135, 206, 235}
\definecolor{reliability}{RGB}{200, 200, 200}
\definecolor{generalizability}{RGB}{180, 180, 180}

\definecolor{attack}{RGB}{255,200,200}
\definecolor{defense}{RGB}{200,255,200}
\definecolor{environment}{RGB}{200,200,255}

\definecolor{lightgray}{HTML}{e0e0e0}

\tikzset{
  forked edges/.style={},
  grey/.style={fill=gray!30},
  extraction-middle/.style={draw=extraction, fill=middle-color!40, text opacity=1, align=center, fill opacity=.5, text=black, font=\scriptsize, inner sep=3pt},
  extraction-leaf/.style={draw=extraction, fill=leaf-color!40, text opacity=1, align=left, fill opacity=.5, text=black, font=\scriptsize, inner sep=3pt},
  inversion-middle/.style={draw=inversion, fill=middle-color!40, text opacity=1, align=center, fill opacity=.5, text=black, font=\scriptsize, inner sep=3pt},
  inversion-leaf/.style={draw=inversion, fill=leaf-color!40, text opacity=1, align=left, fill opacity=.5, text=black, font=\scriptsize, inner sep=3pt},
  others-middle/.style={draw=others, fill=middle-color!40, text opacity=1, align=center, fill opacity=.5, text=black, font=\scriptsize, inner sep=3pt},
  others-leaf/.style={draw=others, fill=leaf-color!40, text opacity=1, align=left, fill opacity=.5, text=black, font=\scriptsize, inner sep=3pt},
  reliability-middle/.style={draw=reliability, fill=middle-color!40, text opacity=1, align=center, fill opacity=.5, text=black, font=\scriptsize, inner sep=3pt},
  reliability-leaf/.style={draw=reliability, fill=leaf-color!40, text opacity=1, align=left, fill opacity=.5, text=black, font=\scriptsize, inner sep=3pt},
  generalizability-middle/.style={draw=generalizability, fill=middle-color!40, text opacity=1, align=center, fill opacity=.5, text=black, font=\scriptsize, inner sep=3pt},
  generalizability-leaf/.style={draw=generalizability, fill=leaf-color!40, text opacity=1, align=left, fill opacity=.5, text=black, font=\scriptsize, inner sep=3pt}
}

\forestset{
  default preamble={
    for tree={
      align=center,
      parent anchor=south,
      child anchor=north,
    }
  },
}

\AtBeginDocument{%
  }

\setcopyright{acmlicensed}
\copyrightyear{2025}
\acmYear{2025}
\setcopyright{cc}
\setcctype{by}
\acmConference[KDD '25]{Proceedings of the 31st ACM SIGKDD Conference on Knowledge Discovery and Data Mining V.2}{August 3--7, 2025}{Toronto, ON, Canada}
\acmBooktitle{Proceedings of the 31st ACM SIGKDD Conference on Knowledge Discovery and Data Mining V.2 (KDD '25), August 3--7, 2025, Toronto, ON, Canada}
\acmDOI{10.1145/3711896.3736573}
\acmISBN{979-8-4007-1454-2/2025/08}

\begin{document}

\title{A Survey on Model Extraction Attacks and Defenses for \\Large Language Models}

\author{Kaixiang Zhao}
\email{kzhao5@nd.edu}
\orcid{0009-0005-8174-0581}
\affiliation{%
  \institution{University of Notre Dame}
  \city{South Bend}
  \state{Indiana}
  \country{USA}
}

\author{Lincan Li}
\email{ll24bb@fsu.edu}
\orcid{0000-0003-3797-4055}
\affiliation{%
  \institution{Florida State University}
  \city{Tallahassee}
  \state{Florida}
  \country{USA}
}
\authornote{Co-first author}

\author{Kaize Ding}
\email{kaize.ding@northwestern.edu}
\orcid{0000-0001-6684-6752}
\affiliation{%
  \institution{Northwestern University}
  \city{Evanston}
  \state{Illinois}
  \country{USA}
}

\author{Neil Zhenqiang Gong}
\email{neil.gong@duke.edu}
\orcid{0000-0002-9900-9309}
\affiliation{%
  \institution{Duke University}
  \city{Durham}
  \state{North Carolina}
  \country{USA}
}

\author{Yue Zhao}
\email{yzhao010@usc.edu}
\orcid{0000-0003-3401-4921}
\affiliation{%
  \institution{University of Southern California}
  \city{Los Angeles}
  \state{California}
  \country{USA}
}

\author{Yushun Dong}
\email{yushun.dong@fsu.edu}
\orcid{0000-0001-7504-6159}
\affiliation{%
  \institution{Florida State University}
  \city{Tallahassee}
  \state{Florida}
  \country{USA}
}
\authornote{Corresponding author.}







\renewcommand{\shortauthors}{Kaixiang Zhao et al.}

\begin{abstract}
Model extraction attacks pose significant security threats to deployed language models, potentially compromising intellectual property and user privacy. This survey provides a comprehensive taxonomy of LLM-specific extraction attacks and defenses, categorizing attacks into functionality extraction, training data extraction, and prompt-targeted attacks. We analyze various attack methodologies including API-based knowledge distillation, direct querying, parameter recovery, and prompt stealing techniques that exploit transformer architectures. We then examine defense mechanisms organized into model protection, data privacy protection, and prompt-targeted strategies, evaluating their effectiveness across different deployment scenarios. We propose specialized metrics for evaluating both attack effectiveness and defense performance, addressing the specific challenges of generative language models. Through our analysis, we identify critical limitations in current approaches and propose promising research directions, including integrated attack methodologies and adaptive defense mechanisms that balance security with model utility. This work serves NLP researchers, ML engineers, and security professionals seeking to protect language models in production environments.
\end{abstract}




\maketitle

\section{Introduction}

A growing number of 
large language models (LLMs), often developed with substantial investments by companies like OpenAI and Google, bear significant commercial value and advanced AI capabilities
~\cite{marechal2024flow, sachs2023ai,achiam2023gpt,team2023gemini,guo2025deepseek}. The considerable resources invested and the imperative to protect valuable intellectual property typically restrict the public release of these state-of-the-art models \cite{achiam2023gpt,team2023gemini,guo2025deepseek}. Consequently, these models are commonly made accessible via Machine-Learning-as-a-Service (MLaaS) platforms, which provide API-based query access to pre-trained models and manage the underlying infrastructure \cite{yao2017complexity, ribeiro2015mlaas, kim2018nsml,philipp2020machine,hunt2018chiron}. However, these platforms also pose significant security challenges, primarily stemming from the potential for unauthorized replication of model functionality or theft of underlying intellectual property. Such activities, collectively referred to as model extraction attacks (MEAs)---or model stealing attacks---have emerged as a critical threat to private-owned LLMs offered via these service interfaces \cite{aguilera2025llm, patil2023can, tete2024threat, kesarwani2018model, hsieh2023distilling,wang2025cegacosteffectiveapproachgraphbased,cheng2025misleader}. Recent research demonstrates that private-owned models can be successfully replicated, and their performance on relevant benchmarks even surpassed, through systematic distillation approaches. These methods often require small amounts of training data, and empirical evidence further supports this phenomenon \cite{huang2024o1, qin2024o1, yang2025distilling, hsieh2023distilling, fang2025knowledge, liu2024ddk, lee2025quantificationlargelanguagemodel}. Findings from these studies suggest that many well-known closed-source and even some open-source models exhibit high degrees of distillation from leading commercial models \cite{lee2025quantificationlargelanguagemodel, baninajjar2024verified, datageneralization}. Some replicated models also show response patterns and identity characteristics strikingly similar to those of the original providers \cite{lee2025quantificationlargelanguagemodel, baninajjar2024verified, datageneralization}. This threat is further validated by ongoing allegations that some AI developers are potentially employing MEAs against such private-owned commercial models to build their own powerful models, which reportedly match the performance of leading commercial models at a significantly lower cost \cite{yang2024distillseq,lee2025quantificationlargelanguagemodel, shirgaonkar2024knowledge,guo2025deepseek}
These controversies, even when specific claims are contested, have intensified debate on model extraction against LLMs. Furthermore, the unique characteristics of LLMs, such as their scale, generative capabilities, introduce vulnerabilities that attackers can exploit with increasing sophistication \cite{zhao2025survey,wang2024unique,yao2024survey,wu2024new}. These attacks typically exploit the model's query APIs to replicate functionality without authorization, leveraging techniques such as systematic API queries \cite{krishna2020thieves,birch2023model,xu2025survey,reith2019efficiently}, parameter inference \cite{tramer2016stealing,rakin2022deepsteal}, training data extraction \cite{carlini2021extracting,gerasimenko2024extracting}, and prompt manipulation \cite{sha2024prompt,perez2022ignore,shen2024prompt}.



To tackle these threats, researchers have proposed various defense strategies that achieve protection at different levels.. These include architectural safeguards such as watermarking \cite{he2022cater,liu2024survey,zhang2024remark,zhaocan,zhang2024personamark} and access control mechanisms \cite{li2024translinkguard,wang2025llm,rathod2025privacy}, training-time interventions like robust architecture design \cite{patil2023can,wei2023jailbroken,halawi2024covert}, and inference-time protections including output sanitization \cite{wang2023self,cheng2025atom,ishibashi2023knowledge} and query monitoring \cite{li2024llm,yuan2024wip,yan2024protecting}. Despite these advances, significant challenges in LLM extraction security persist, including balancing model accessibility with security and utility, protecting diverse LLM architectures and deployments, and countering evolving attack strategies \cite{he2022cater,wang2023self,yao2024survey,vijjini2024exploring,dhingra2024protecting,dong2024safeguarding}.


\noindent\textbf{Core Contributions.} This survey makes four key contributions to the domain of LLM extraction security:
First, we provide a clear taxonomy of LLM extraction attacks, detailing transformer-specific vulnerabilities and generative model risks.
Second, we analyze current defense mechanisms, examining architectural, output control, and monitoring strategies to show their strengths and weaknesses.
Third, we examine key security-utility tradeoffs, showing how defenses affect model quality and usability, to help balance security with performance.
Fourth, we propose an evaluation framework with specific metrics to measure attack success and defense strength, suited for generative models.
Finally, this work identifies research gaps and suggests future directions in this field.

\noindent\textbf{Intended Audience.} This survey targets three primary audiences: (1) NLP researchers investigating security vulnerabilities and protections for language models; (2) ML engineers and practitioners responsible for deploying and securing LLMs in production environments; and (3) Security professionals tasked with protecting organizational AI assets and intellectual property. 
By bridging theoretical foundations with practical implementation considerations, this work provides actionable insights for both academic research and industrial applications in LLM security.
\begin{figure}[t] 
    \centering
    \includegraphics[width=\columnwidth]{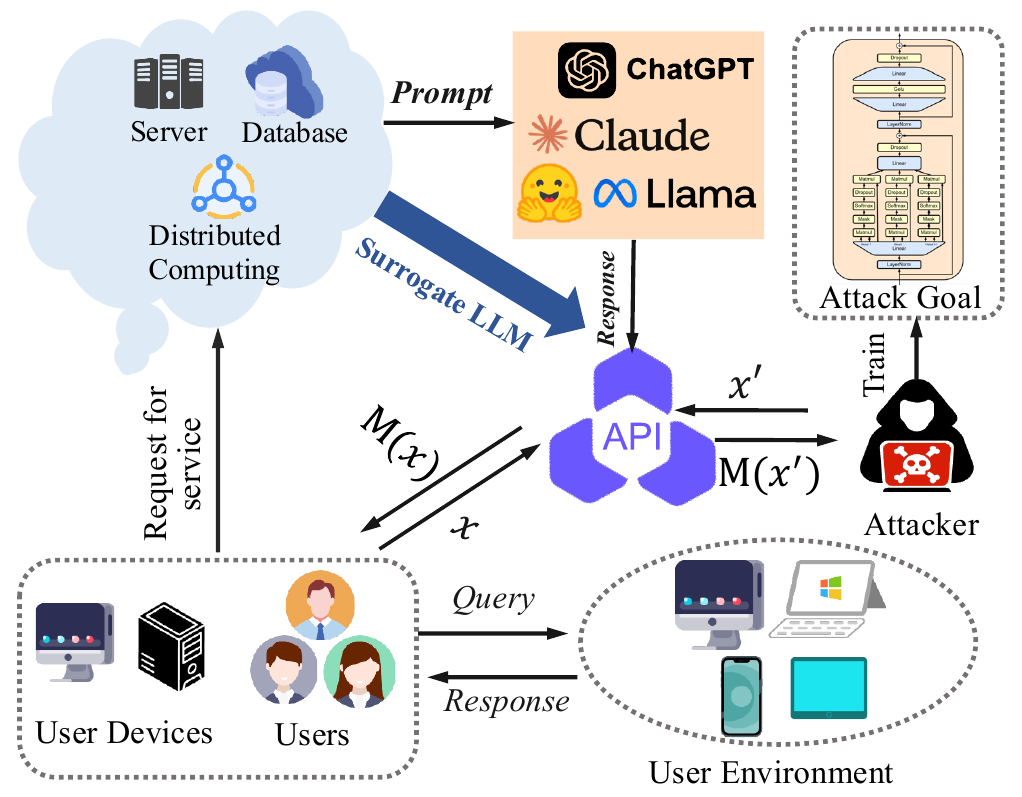}
    \caption{Illustration of the Model Extraction Attack pipeline on Large Language Models.}
    \label{fig:MEA_LLM_diagram} 
\end{figure}


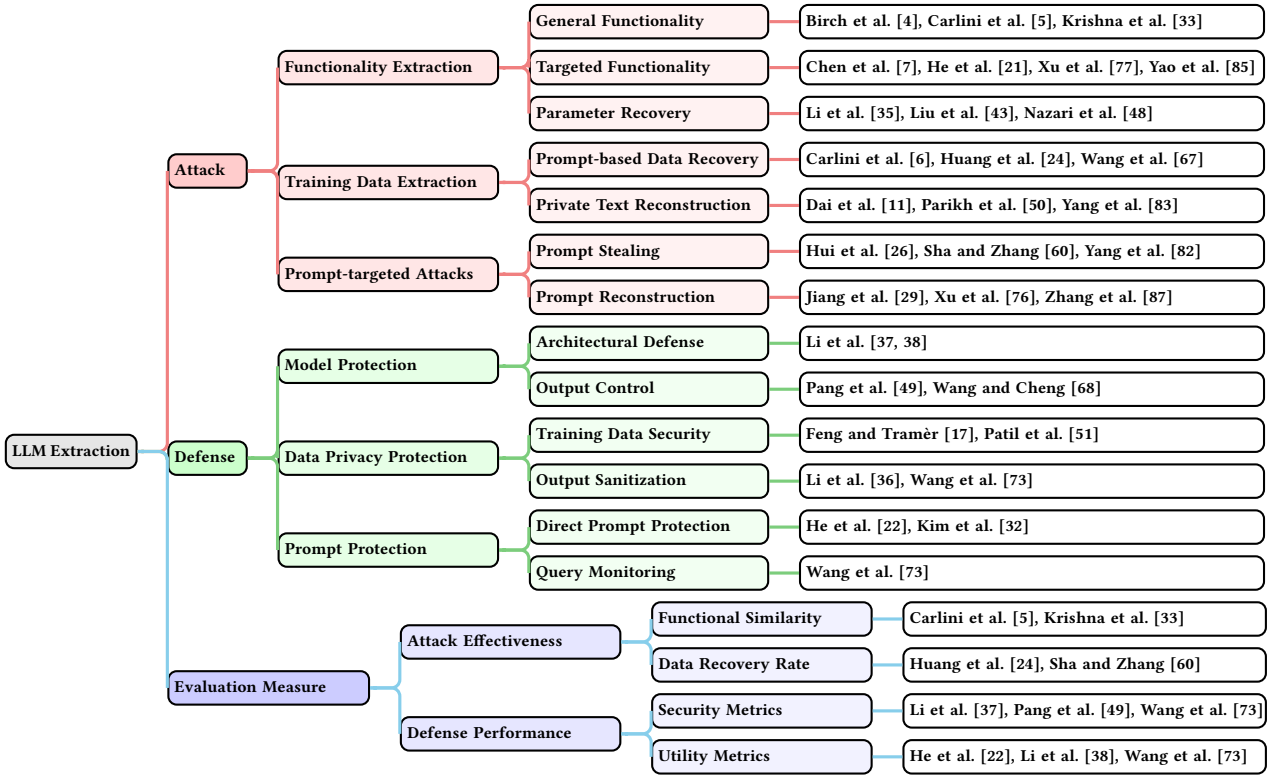
\begin{figure*}[ht]
\centering
\hspace*{-1.2cm}
\resizebox{0.95\textwidth}{!}{
\begin{forest}
  for tree={
    forked edges,
    edge={-, draw=black, line width=1.5pt},
    edge path={
      \noexpand\path[\forestoption{edge}, rounded corners]
      (!u.parent anchor) -- +(-1pt,0pt) -| ([xshift=0pt].child anchor)\forestoption{edge label};
    },
    grow=east,
    reversed,
    anchor=base west,
    parent anchor=east,
    child anchor=west,
    base=middle,
    font=\small\bfseries,
    rectangle,
    draw=black,
    line width=1pt,
    rounded corners,    
    align=left,
    minimum width=2em,
    s sep=5pt,
    l sep=0.5cm,
    inner sep=3pt,
  },
  where level=1{text width=4.5em}{},
  where level=2{text width=11em}{},
  where level=3{text width=11em}{},
  where level=4{edge path={\noexpand\path[\forestoption{edge}] (!u.parent anchor) -- (.child anchor)\forestoption{edge label};}}{},
  where level=5{}{},
  [\textbf{LLM Extraction}, fill=gray!20, align=center, anchor=north, edge=extraction, 
    [Attack, fill=red!20, edge=extraction, text width=3.5em
      [Functionality Extraction, fill=red!10, edge=extraction
        [General Functionality, fill=red!5, text width=12em, edge=extraction
          [\citet{krishna2020thieves,carlini2024stealing,birch2023model}, fill=white, text width=24em, edge=extraction]
        ]
        [Targeted Functionality, fill=red!5, text width=12em, edge=extraction
          [\citet{yao2017complexity,xu2021student,he2021model,chen2021killing}, fill=white, text width=24em, edge=extraction]
        ]
        [Parameter Recovery, fill=red!5, text width=12em, edge=extraction
          [\citet{li2023theoretical,nazari2024llm,liu2022stolenencoder}, fill=white, text width=24em, edge=extraction]
        ]
      ]
      [Training Data Extraction, fill=red!10, edge=extraction
        [Prompt-based Data Recovery, fill=red!5, text width=12em, edge=extraction
          [\citet{carlini2021extracting,huang2022are,wang2024pandora}, fill=white, text width=24em, edge=extraction]
        ]
        [Private Text Reconstruction, fill=red!5, text width=12em, edge=extraction
          [\citet{yang2024unveiling,parikh2022canary,dai2025stealing}, fill=white, text width=24em, edge=extraction]
        ]
      ]
      [Prompt-targeted Attacks, fill=red!10, edge=extraction
        [Prompt Stealing, fill=red!5, text width=12em, edge=extraction
          [\citet{sha2024prompt,yang2024prsa,hui2024pleak}, fill=white, text width=24em, edge=extraction]
        ]
        [Prompt Reconstruction, fill=red!5, text width=12em, edge=extraction
          [\citet{zhang2024extracting,xu2024instructional,jiang2025mimicking}, fill=white, text width=24em, edge=extraction]
        ]
      ]
    ]
    [Defense, fill=green!20, edge=inversion, text width=3.5em
      [Model Protection, fill=green!10, edge=inversion
        [Architectural Defense, fill=green!5, text width=12em, edge=inversion
          [\citet{li2024translinkguard,li2024coreguard}, fill=white, text width=24em, edge=inversion]
        ]
        [Output Control, fill=green!5, text width=12em, edge=inversion
          [\citet{wang2024guardemb,pang2025modelshield}, fill=white, text width=24em, edge=inversion]
        ]
      ]
      [Data Privacy Protection, fill=green!10, edge=inversion
        [Training Data Security, fill=green!5, text width=12em, edge=inversion
          [\citet{patil2023can,feng2024privacy}, fill=white, text width=24em, edge=inversion]
        ]
        [Output Sanitization, fill=green!5, text width=12em, edge=inversion
          [\citet{wang2023self,li2024llm}, fill=white, text width=24em, edge=inversion]
        ]
      ]
      [Prompt Protection, fill=green!10, edge=inversion
        [Direct Prompt Protection, fill=green!5, text width=12em, edge=inversion
          [\citet{he2022cater,kim2024protection}, fill=white, text width=24em, edge=inversion]
        ]
        [Query Monitoring, fill=green!5, text width=12em, edge=inversion
          [\citet{wang2023self}, fill=white, text width=24em, edge=inversion]
        ]
      ]
    ]
    [Evaluation Measure, fill=blue!20, edge=others, text width=10em
      [Attack Effectiveness, fill=blue!10, edge=others, text width=11em,
        [Functional Similarity, fill=blue!5, text width=11em, edge=others
          [\citet{carlini2024stealing,krishna2020thieves}, fill=white, text width=18.6em, edge=others, edge path={\noexpand\path[\forestoption{edge}] (!u.parent anchor) -- (.child anchor)\forestoption{edge label};}]
        ]
        [Data Recovery Rate, fill=blue!5, text width=11em, edge=others
          [\citet{huang2022are,sha2024prompt}, fill=white, text width=18.6em, edge=others, edge path={\noexpand\path[\forestoption{edge}] (!u.parent anchor) -- (.child anchor)\forestoption{edge label};}]
        ]
      ]
      [Defense Performance, fill=blue!10, edge=others, text width=11em,
        [Security Metrics, fill=blue!5, text width=11em, edge=others
          [\citet{li2024translinkguard,wang2023self,pang2025modelshield}, fill=white, text width=18.6em, edge=others, edge path={\noexpand\path[\forestoption{edge}] (!u.parent anchor) -- (.child anchor)\forestoption{edge label};}]
        ]
        [Utility Metrics, fill=blue!5, text width=11em, edge=others
          [\citet{li2024coreguard,he2022cater,wang2023self}, fill=white, text width=18.6em, edge=others, edge path={\noexpand\path[\forestoption{edge}] (!u.parent anchor) -- (.child anchor)\forestoption{edge label};}]
        ]
      ]
    ]
  ]
\end{forest}
}
\vspace{-0mm}
\caption{A Taxonomy of Model Extraction Attacks \& Defenses on Large Language Models.}
\vspace{-3mm}
\label{fig:taxonomy-techniques}
\end{figure*}

\section{Preliminaries}

\subsection{Model Extraction Basics}

Model Extraction Attacks (MEAs) on Large Language Models (LLMs) aim to illicitly replicate the functionality, knowledge, or parameters of a target model. We first formally define the threat model.

\begin{myDef1}[Threat Model]
\label{def:llm_extraction_threat_model} 

Given a target private-owned LLM \(M_T\), black-box query access (allowing submission of input prompts \(x\) to receive outputs \(y = M_T(x)\) without internal model access), and a potential query budget \(B\). The attacker aims to train a surrogate model \(M_S\) that replicates the input-output functional pattern of \(M_T\).
\end{myDef1}



Specifically, in the LLM context, the target model $M$ typically processes input text $x$ (often structured as a prompt) and generates output text $y = M(x)$. An adversary with query access to $M$ constructs an extracted dataset $D_{ext} = \{(x_i, M(x_i)) | x_i \sim X, 1 \leq i \leq N\}$, where $X$ represents the input space of possible prompts and $N$ is the number of queries. Using this dataset, the adversary trains a surrogate model $M'$ that approximates the behavior of $M$ by minimizing a loss function
$M' = \arg\min_{M' \in H} \sum_{(x,y) \in D_{ext}} \mathcal{L}(M'(x), y),$
where $H$ represents the hypothesis space of possible models and $\mathcal{L}$ measures the discrepancy between outputs. 

\subsection{Extraction Attack Process}

The LLM extraction process involves three key phases: query generation, response collection, and surrogate model training. In query generation, an adversary, operating from their \texttt{User Environment} as shown in Figure~\ref{fig:MEA_LLM_diagram}, designs prompts (\(x'\)) that effectively probe the model's capabilities. During response collection, these queries are submitted to the target model through its API service, and the corresponding outputs (\(M(x')\)) are recorded for training data.

The surrogate model training phase leverages the collected input-output pairs (\(x', M(x')\)) to train a model that mimics the target LLM. This may involve fine-tuning a smaller pre-trained model, distilling knowledge into a more compact architecture, or training from scratch. The training process must address challenges specific to language modeling, including handling variable-length sequences and replicating the specific characteristics of the target model. As illustrated in Figure~\ref{fig:MEA_LLM_diagram}, the adversary's ultimate goal is to develop a surrogate model, \(M'(\cdot)\), that approximates the target model, \(M(\cdot)\).

\section{Taxonomy}

Figure~\ref{fig:taxonomy-techniques} presents our taxonomy of model extraction attacks and defenses for LLMs. We categorize MEA attacks against LLMs into three categories: functionality extraction targeting model behavior replication, training data extraction recovering training examples, and prompt-targeted attacks stealing valuable prompts. Defense mechanisms are similarly organized: model protection strategies secure architecture and outputs, data privacy protection safeguards training data, and prompt protection protects private-owned prompts and monitors queries. For evaluation, we identify metrics covering attack effectiveness and defense performance. 

\section{Model Extraction Attacks}

Model extraction attacks against LLMs present significant threats to intellectual property, and user privacy. This section details three primary attack types: \textit{Model Functionality Extraction}, \textit{Training Data Extraction}, and \textit{Prompt-targeted Attacks.}

\subsection{Model Functionality Extraction}
Model functionality extraction attacks aim to replicate a target LLM's capabilities through black-box query interactions. Given a target language model $M: X \rightarrow Y$, the adversary collects a dataset $D_{ext} = \{(x_i, M(x_i)) | x_i \sim X, 1 \leq i \leq N\}$ through $N$ queries, constrained by budget $B$. They then train a surrogate model
\begin{equation}
M' = \arg\min_{M' \in \mathcal{H}} \sum_{(x,y) \in D_{ext}} \mathcal{L}(M'(x), y),
\end{equation}
where $\mathcal{L}$ measures output discrepancy. 
Model functionality extraction attacks are categorized into three types: \textit{general functionality extraction}, \textit{targeted functionality extraction}, and \textit{parameter/architecture recovery}. General functionality extraction systematically transfers the overall functionality of a target LLM by querying it with a broad and diverse set of inputs to create a comprehensive dataset of input-output pairs. This dataset is then used to train a surrogate LLM that replicates the target LLM's behavior. In contrast, targeted functionality extraction focuses on crafting specific prompts to extract particular behaviors or capabilities from the model, often targeting narrow functionalities or specific response patterns. Parameter/architecture recovery differs from both by attempting to reverse-engineer internal components, such as weights.

\subsubsection{General Functionality Extraction}

General functionality extraction attacks leverage authorized access to target LLMs through their public interfaces to systematically transfer knowledge to an attacker-controlled model. Unlike traditional knowledge distillation requiring direct access to model parameters, these attacks operate purely through input-output interactions. \citet{krishna2020thieves} demonstrated the feasibility of model extraction against BERT-based APIs, which required only query access to create functional clones with competitive performance. Their approach revealed a fundamental vulnerability: as commercial models standardize on transformer architectures, the technical barriers to model extraction diminish dramatically. \citet{birch2023model} proposed a model leeching attack specifically targeting LLMs that achieved high functional similarity with limited computational resources. Most concerning, \citet{carlini2024stealing} successfully demonstrated extraction attacks against production-scale language models, which showed that even billion-parameter commercial systems remain vulnerable. Their approach requires no insider access, which highlights a widening asymmetry between the immense resources required to develop frontier models and the relatively modest investment needed to steal their capabilities. The core insight is that knowledge distillation attacks become practical as the architecture gap between commercial models narrows.

\subsubsection{Targeted Functionality Extraction}

Targeted functionality extraction attacks use carefully designed prompts to efficiently extract model capabilities by analyzing the model's responses. \citet{yao2017complexity} established early foundations by analyzing vulnerabilities in MLaaS systems. Their work showed that significant model functionality could be extracted even with limited queries. This vulnerability has increased with modern LLMs, as demonstrated by \citet{krishna2020thieves}, who showed that strategic querying can rapidly approximate the functionality of BERT-based models. \citet{he2021model} further proved that model functionality can be extracted, and the resulting clone exhibits similar vulnerability to adversarial examples. This suggests that fundamental behavioral characteristics transfer during extraction. \citet{chen2021killing} expanded this research by demonstrating that BERT-based APIs are vulnerable to dual-purpose attacks, which simultaneously extract model functionality and user attributes. These findings compound the security implications of extraction attacks. Modern API-based attacks are distinguished by their increasing query efficiency. Early techniques required millions of queries, but newer methods use information-theoretic approaches to identify high-value prompts that reveal maximum model behavior with minimal interaction. \citet{xu2021student} highlighted this efficiency trend with their imitation attack, which enables students to surpass teachers through optimized query strategies. This evolution presents a concerning trend. As attack efficiency improves, detection becomes more difficult, creating a security gap where model functionality can be extracted before defenses are engaged.

\subsubsection{Parameter/Architecture Recovery}

Parameter and architecture recovery attacks are among the most technically challenging forms of model extraction. These attacks aim to reverse-engineer specific model components, such as weights or structural designs, rather than replicate behavior. They are particularly concerning for edge-deployed models, such as those hosted on local devices like smartphones or IoT hardware, where adversaries may have physical access. \citet{li2023theoretical} provided theoretical foundations for gradient leakage in transformer architectures. Their work demonstrates that attention mechanisms leak more information than traditional neural networks. \citet{nazari2024llm} exploited this vulnerability in practical attacks. They presented a fingerprinting methodology that reveals critical architectural details of edge-deployed LLMs. \citet{liu2022stolenencoder} showed that even self-supervised learning encoders can be stolen, which compromises the foundational components of modern language models. Transformer-based LLMs, despite their complexity, exhibit structure-revealing patterns during inference. Self-attention mechanisms expose architectural information through output behavior, which can be systematically probed and analyzed. Complete parameter recovery remains impractical for billion-parameter models. However, partial recovery of key components creates vulnerabilities that enable targeted attacks or provide competitive advantages. This vulnerability increases in distributed computing environments, where adversaries can exploit expanded access to model components, as highlighted by \citet{zhao2025survey}.

\subsection{Training Data Extraction} \label{training_data_extraction}
Training data extraction attacks aim to recover specific data points from a model's training dataset, which exploits the model's tendency to memorize and reproduce examples from its training data \cite{carlini2021extracting}. These attacks are considered a form of model extraction because recovering the training data constitutes a direct extraction of potentially sensitive and private information from the model. The training data is the source of the model's learned patterns and behaviors. Given a model $M$ trained on dataset $D_{train}$, an adversary crafts prompts $P = \{p_1,...,p_N\}$ with budget $|P| \leq B$ to extract training data
\begin{equation}
E(M) = \{d \in D_{train} : \exists p \in P \text{ s.t. } \text{sim}(M(p), d) > \tau\},
\end{equation}
where $\text{sim}(\cdot,\cdot)$ measures similarity and $\tau$ is a threshold.
We categorize training data extraction attacks into two types: \textit{prompt-based} \textit{data recovery} and \textit{private text reconstruction}. Prompt-based data recovery focuses on retrieving memorized training examples using carefully crafted prompts, which target specific data patterns such as PII or rare text formats. In contrast, private text reconstruction goes beyond verbatim extraction by inferring and reconstructing sensitive information through techniques like activation inversion and canary extraction.

\subsubsection{Prompt-based Data Recovery}

Prompt-based data recovery attacks exploit LLMs' tendency to memorize training examples, which allows specific prompts to retrieve memorized data. \citet{carlini2021extracting} demonstrated the practical feasibility of extracting exact training examples from GPT-2, which showed that models memorize and can be induced to reproduce specific sequences from their training data. Their work established that the success of such extractions increases with model size and complexity. This indicates that more advanced models may unintentionally increase privacy risks. \citet{zhang2023ethicist} refined this approach with their ETHICIST system, which uses loss-smoothed soft prompting and calibrated confidence estimation to improve the efficiency of targeting specific types of training data. \citet{huang2022are} specifically investigated personal information leakage, which revealed that pre-trained LLMs frequently memorize and expose personally identifiable information (PII) when properly prompted. Their findings suggest that uniquely formatted data patterns, such as email addresses and phone numbers, are particularly vulnerable to extraction. This creates a direct privacy risk for individuals whose information appeared in the training data. \citet{wang2024pandora} further reinforced this vulnerability by developing more precise extraction techniques that can distinguish between memorized and generated content. Average memorization rates may appear low in aggregate measurements. However, specific types of data, such as unique, rare, or unusually formatted text, are disproportionately memorized and more vulnerable to extraction. This creates a practical privacy risk with significant implications for data protection and regulatory compliance.

\subsubsection{Private Text Reconstruction}

Private text reconstruction attacks go beyond extracting verbatim examples, which recover sensitive information through inference and reconstruction techniques. \citet{zhang2022text} demonstrated ``Text Revealer," which successfully reconstructs private text from transformer models by exploiting their self-attention mechanisms. This work showed that internal model states contain recoverable private information. \citet{yang2024unveiling} extended this concept to code models, which revealed that specialized domains exhibit distinct patterns for memorization that can be exploited for precise extraction. \citet{parikh2022canary} introduced canary extraction, which showed that models memorize and leak specific inserted data patterns at rates that correlate with training exposure. This technique evaluates memorization risks by analyzing how easily models reproduce strategically inserted markers. \citet{dai2025stealing} demonstrated that activation inversion attacks can reconstruct training data in decentralized training environments, which highlights how internal model representations can be exploited to recover private information. Early research demonstrated that extraction was possible in certain scenarios. Newer techniques employ principled approaches, which target specific data types with high precision. This evolution suggests that privacy protections must advance beyond preventing obvious memorization to address subtle statistical patterns that enable reconstruction. As \citet{gerasimenko2024extracting} argue in their analysis of extraction risks, these threats require reconsideration of how language models are deployed, particularly in domains handling sensitive information.

\subsection{Prompt-targeted Attacks}
Prompt-targeted attacks target valuable prompts used to guide LLM behavior. Given a model $M$ and a private-owned prompt $P^*$, an adversary with query access to $M(P^*,x_i)$ for inputs $x_i$ aims to reconstruct a similar prompt
\begin{equation}
\hat{P} = \arg\max_{P} \{\text{sim}(P, P^*) : \text{sim}(M(P,x), M(P^*,x)) > \tau, \forall x \in X_{test}\},
\end{equation}
where $\text{sim}(\cdot,\cdot)$ measures similarity and $X_{test}$ is a validation set.
We categorize prompt-targeted attacks into two types: \textit{prompt stealing} and \textit{prompt reconstruction}. Prompt stealing focuses on extracting carefully engineered prompts, such as system instructions, which represent significant commercial assets. In contrast, prompt reconstruction aims to recover instruction patterns and few-shot examples by analyzing model outputs, which retain traces of their prompting context.
\subsubsection{Prompt Stealing}

Prompt stealing attacks target the intellectual property embedded in carefully engineered prompts, which represent significant commercial assets in LLM applications. \citet{sha2024prompt} pioneered research into systematic prompt stealing, which demonstrated that prompts could be extracted through strategic interactions with LLM applications. Their work showed that as models become more capable, the value increasingly shifts to prompting strategies, which creates new intellectual property vulnerabilities. \citet{yang2024prsa} extended this work with PRSA (Prompt Stealing Attacks), which showed that even complex system prompts could be reconstructed with limited interaction by exploiting model responses to carefully designed queries. \citet{hui2024pleak} demonstrated PLeak attacks, which reveal that commercial applications using LLMs frequently leak their underlying prompts through inconsistent output filtering. These vulnerabilities can be exploited even in production systems. \citet{liang2024my} investigated the fundamental mechanisms of prompt leakage in customized LLMs, which identified architectural vulnerabilities that enable extraction. Their work revealed that prompts encoded as system instructions leave detectable patterns in model responses, which facilitate reconstruction. Prompt stealing represents an economic threat rather than merely a technical vulnerability. Businesses increasingly invest in prompt engineering to differentiate their AI offerings. Prompt theft provides a low-cost mechanism to appropriate competitive advantages. This dynamic creates a strategic tension where the most valuable prompts also become the most attractive targets for extraction. This may undermine incentives for innovation in prompt engineering and specialized model development.

\subsubsection{Prompt Reconstruction}

Prompt reconstruction attacks focus on recovering the instruction patterns and few-shot examples, which are used to specialize models for specific tasks and potentially compromise both intellectual property and security boundaries. \citet{zhang2024extracting} demonstrated techniques for inverting LLM outputs to extract the prompts that generated them, which showed that responses contain sufficient information to reconstruct significant portions of input prompts. This work revealed that model outputs inherently retain traces of their prompting context, which creates fundamental information leakage. \citet{xu2024instructional} introduced instructional fingerprinting of LLMs, which revealed that models retain detectable patterns from their instruction tuning that can be analyzed to reconstruct training processes. This research highlights how instruction tuning creates behavioral patterns that persist through inference. These patterns can potentially compromise the privacy of tuning data. \citet{jiang2025mimicking} presented an advanced approach for dynamic command generation in LLM tool-learning systems, which showed that instruction patterns can be systematically extracted and exploited. \citet{mehrotra2024tree} further developed this concept with their ``Tree of Attacks" methodology, which systematically explores variations in prompt space to reconstruct effective instruction patterns. Prompt reconstruction research reveals that LLMs unintentionally encode aspects of their prompting history into their responses. This creates an information leakage channel where careful analysis of outputs can reveal the prompting techniques used to elicit them. This vulnerability undermines attempts to create secure boundaries between system instructions and user-facing functionality.It poses risks to the intellectual property of commercial prompts and the security measures designed to protect them. As LLMs increasingly rely on sophisticated prompting for alignment and safety, this vulnerability raises concerns about the robustness of current security approaches and the potential for adversaries to systematically compromise key model safety features.

\section{Model Extraction Defenses}
Defending against model extraction attacks requires a approach that addresses vulnerabilities at different levels of model deployment and operation. This section examines defense strategies
organized into three categories: \textit{Model Protection}, \textit{Data Privacy Protection}, and \textit{Prompt Protection}.

\subsection{Model Protection}
Model protection implements defensive mechanisms to prevent unauthorized model extraction or functionality replication. Given a model $M$, these defenses aim to create a protected version $M'$ that maximizes legitimate utility while minimizing extraction success
\begin{equation}
M' = \arg\max_{M' \in \mathcal{M}} \{U(M', X_{leg}) - \lambda E(M', X_{adv})\},
\end{equation}
where $U$ measures utility for legitimate inputs $X_{leg}$, $E$ measures extraction success for adversarial inputs $X_{adv}$, and $\lambda$ balances the trade-off.
We categorize model protection mechanisms into two types: \textit{architectural defenses} and \textit{output control}. Architectural defenses modify the LLM's internal structure, such as embedding watermarks or altering attention mechanisms, to prevent unauthorized extraction. In contrast, output control strategies manipulate LLM responses to reduce extraction risks without altering the underlying architecture.
\subsubsection{Architectural Defense}

Architectural defense mechanisms integrate security features directly into the model's structure, which prevents unauthorized extraction. \citet{li2024translinkguard} proposed TransLinkGuard, which safeguards transformer models in edge deployments by manipulating attention mechanisms to embed watermarks, which resist extraction and maintain performance. This approach demonstrates that architectural defenses can be implemented with minimal computational overhead, which is critical for resource-constrained edge devices. \citet{li2024coreguard} extended this concept with CoreGuard, which protects foundational capabilities through structural modifications that make extraction attempts produce degraded model clones. The key insight from these approaches is that effective architectural defenses must target the specific mechanisms that are exploited during extraction rather than applying general security principles. Transformer-based models present unique vulnerabilities because their self-attention mechanisms inadvertently expose architectural information during inference. By strategically altering these attention patterns, defenders can create asymmetric advantages, which allow legitimate users to experience minimal performance impact while extractors receive functionally compromised knowledge. However, these approaches often require significant modifications to existing model architectures. This limits their applicability to newly developed models and makes retrofitting protection onto existing deployments challenging. Consequently, such defenses are often best suited for new models where security can be incorporated from the initial design phase.

\subsubsection{Output Control}

Strategies to control model responses focus on reducing the risk of unauthorized data extraction by manipulating outputs, which avoids modifying the underlying architecture. \citet{wang2024guardemb} introduced GuardEmb, which implements watermarking techniques that adapt to usage for embedding services. This technique alters response patterns in ways that are detectable by service providers but difficult for attackers to avoid. \citet{pang2025modelshield} developed ModelShield, which uses watermarking methods that adjust dynamically to provide robust protection against extraction attacks. This approach strategically perturbs responses while maintaining high utility for legitimate users. These techniques demonstrate that controlled output manipulation can create barriers to extraction with minimal performance impact. The fundamental insight is that strategies to control model responses offer deployment flexibility compared to architectural approaches, which allows them to be implemented as external layers added to the model's API without modifying model internals. This advantage enables retrofitting protection onto existing deployed models and allows for dynamic adjustment of protection levels based on threat assessments. This is particularly valuable for securing proprietary, closed-source models where internal modification is not an option. The challenge remains in balancing the degree of response alteration. Insufficient changes fail to prevent extraction, while excessive modification compromises legitimate user experience. As extraction techniques grow more sophisticated, strategies to control model responses must evolve into adaptive methods, which dynamically calibrate protection based on usage patterns and potential extraction signals.

\subsection{Data Privacy Protection}
Data privacy protection mechanisms aim to prevent extraction of private information from language models. Similarly as we mentioned in the Section \ref{training_data_extraction}, data privacy is also considered as part of model protection \cite{carlini2021extracting}. Given a model $M$ trained on dataset $D$ containing private information $P \subset D$, these defenses create a protected model $M'$ that minimizes privacy leakage while preserving utility:
\begin{equation}
M' = \arg\min_{M' \in \mathcal{M}} \{L(M', P) + \lambda \mathcal{D}(M', M)\},
\end{equation}
where $L$ measures privacy leakage, $\mathcal{D}$ measures model utility deviation, and $\lambda$ balances the trade-off. We categorize data privacy protection mechanisms into two types: \textit{training data security} and \textit{output sanitization}. Training data security focuses on preventing sensitive information from being memorized during training or removing it after exposure, using techniques like differential privacy or selective knowledge removal. In contrast, output sanitization modifies or filters LLM responses to prevent private information leakage during inference.
\subsubsection{Training Data Security}

Methods to secure training data aim to prevent the extraction of sensitive information, which is embedded in model parameters during training. \citet{patil2023can} investigated whether sensitive information could be deleted from LLMs after training. They evaluated various defense objectives against extraction attacks and found that removing specific knowledge can mitigate vulnerabilities while preserving general model capabilities. Their work demonstrates that protecting training data requires preventive measures during training, which must be complemented by corrective measures after potential exposure. The key insight is that securing training data increasingly requires targeted protection of specific information types. Blanket protection approaches are often less effective. While traditional differential privacy techniques apply uniform noise to all training signals, emerging research shows that protecting specific information categories achieves better results in balancing privacy and utility. However, the fundamental challenge remains: models inherently memorize training examples as part of their learning process. This makes complete prevention of training data extraction theoretically difficult. This creates an ongoing tension between learning effectiveness and privacy protection, which drives research into novel training methods that limit memorization while maintaining learning capacity.

\subsubsection{Output Sanitization}

Output sanitization focuses on modifying or filtering model responses, which prevents the leakage of private information. \citet{wang2023self} proposed Self-Guard, which empowers language models to detect and sanitize their own outputs. It uses self-monitoring mechanisms, which filter potentially sensitive information before returning responses. This approach demonstrates that models can be trained to recognize and suppress private information in their own generations, which creates an additional protection layer without external filtering systems. The critical insight is that output sanitization operates on a fundamentally different principle than simple content blocking. Rather than applying static rules, effective sanitization must adapt filtering based on contextual information sensitivity. This requires models to develop the ability to understand the implications of generated information and its privacy risks, which represents a higher-order capability beyond basic content generation. The challenge remains in developing sanitization approaches, which maintain content coherence and utility while removing sensitive elements. This is particularly difficult in contexts where private information forms the central subject of legitimate queries.

\subsection{Prompt Protection}
Prompt protection mechanisms safeguard private prompts and instruction patterns, which represent valuable intellectual property. Given a prompt \(P\) that an organization wishes to protect, these defenses implement safeguards, which maximize detection of unauthorized use and minimize impact on legitimate functionality:
\begin{equation}
\arg\max_{D \in \mathcal{D}} \{\text{TPR}(D, P, X_{adv}) - \lambda \text{Impact}(D, P, X_{leg})\},
\end{equation}
where $\text{TPR}$ measures true positive rate of detecting unauthorized prompt use from adversarial queries $X_{adv}$, $\text{Impact}$ measures the effect on legitimate queries $X_{leg}$, and $\lambda$ balances security and utility.
We categorize prompt protection mechanisms into two types: \textit{direct prompt protection} and \textit{query monitoring}. Direct prompt protection focuses on safeguarding private prompts and instruction patterns through techniques like watermarking and obfuscation, which prevent unauthorized use and detect derivative works. In contrast, query monitoring systems analyze user interactions to identify suspicious behaviors indicative of extraction attempts.
\subsubsection{Direct Prompt Protection}

This mechanism protects private prompts and instruction patterns, which represent valuable intellectual property. \citet{he2022cater} introduced CATER, a conditional watermarking system, which protects intellectual property in text generation APIs by embedding subtle markers that identify prompts and maintain generation quality. \citet{kim2024protection} proposed detailed protection strategies for LLM environments, which use prompt protection techniques to prevent unauthorized access to system instructions. These approaches demonstrate that prompt protection must balance detection capabilities with generation quality to ensure effective protection without degrading user experience. The key insight is that prompt protection has become an emerging priority in the LLM ecosystem as commercial value increasingly shifts from model weights to prompting techniques. As businesses invest substantial resources in developing specialized prompts for specific applications, the economic incentive for prompt theft grows proportionally. Protection mechanisms must therefore prevent direct copying. They must also detect derivative works, which preserve the functionality of the original prompts while modifying their wording, structure, or formatting. This creates a conceptual parallel to traditional software protection, where both code and functionality require safeguarding against unauthorized reproduction.

\subsubsection{Query Monitoring}

Systems to monitor user queries detect and prevent extraction attempts by analyzing user queries and behaviors, which reveal suspicious activity. \citet{wang2023self} demonstrated that self-monitoring approaches can detect suspicious query patterns, which indicate extraction attempts and enable early defensive actions. These systems work by identifying unexpected patterns in user queries, which may suggest extraction attempts, or by recognizing known behaviors associated with extraction attacks. The key insight is that extraction attacks often show distinct behaviors that differ from legitimate usage. These behaviors include systematically exploring model boundaries, using unusual input distributions, or repeatedly testing specific capabilities. Query monitoring establishes typical patterns of legitimate user behavior, which allow it to flag potential extraction attempts for further investigation or response. However, the challenge lies in distinguishing sophisticated extraction attempts from legitimate but unusual usage. This becomes particularly difficult as attackers adapt their techniques to mimic normal user behavior. This creates an ongoing dynamic, which requires monitoring systems to continuously evolve to detect increasingly subtle extraction attempts.

\section{Evaluation Metrics}
Evaluating model extraction attacks and defenses requires specific evaluation criteria, which measure both extraction success and defense effectiveness. Table \ref{tab:defense-effectiveness} summarizes the effectiveness of defense mechanisms against various attack types.
\begin{table*}[!t]
\centering
\small
\setlength{\tabcolsep}{3pt}
\begin{threeparttable}
\caption{Effectiveness of Defense Mechanisms Against Different Model Extraction Attack Types.}
\label{tab:defense-effectiveness}
\begin{tabular}{c|c|c|c|c|c|c|c}
\toprule
\multirow{2}{*}{\textbf{Defense Mechanism}} & \multicolumn{3}{c|}{\textbf{Functionality Extraction}} & \multicolumn{2}{c|}{\textbf{Training Data Extraction}} & \multicolumn{2}{c}{\textbf{Prompt-targeted Attacks}} \\
\cline{2-8}
 & \textbf{General} & \textbf{Targeted} & \textbf{Parameter} & \textbf{Prompt-targeted} & \textbf{Private Text} & \textbf{Prompt} & \textbf{Prompt} \\
 &  \textbf{Functionality} & \textbf{Functionality} & \textbf{Recovery} & \textbf{Recovery} & \textbf{Reconstruction} & \textbf{Stealing} & \textbf{Reconstruction} \\
\midrule
\textbf{Architectural Defense [1]} & \cellcolor{green!25}High & \cellcolor{green!15}Medium & \cellcolor{green!25}High & \cellcolor{yellow!15}Low & \cellcolor{yellow!15}Low & \cellcolor{gray!15}Minimal & \cellcolor{gray!15}Minimal \\
\hline
\textbf{Output Control [2]} & \cellcolor{green!25}High & \cellcolor{green!25}High & \cellcolor{yellow!15}Low & \cellcolor{green!15}Medium & \cellcolor{green!15}Medium & \cellcolor{yellow!15}Low & \cellcolor{yellow!15}Low \\
\hline
\textbf{Training Data Security [3]} & \cellcolor{yellow!15}Low & \cellcolor{gray!15}Minimal & \cellcolor{gray!15}Minimal & \cellcolor{green!25}High & \cellcolor{green!25}High & \cellcolor{gray!15}Minimal & \cellcolor{gray!15}Minimal \\
\hline
\textbf{Output Sanitization [4]} & \cellcolor{yellow!15}Low & \cellcolor{yellow!15}Low & \cellcolor{gray!15}Minimal & \cellcolor{green!25}High & \cellcolor{green!25}High & \cellcolor{yellow!15}Low & \cellcolor{yellow!15}Low \\
\hline
\textbf{Prompt Protection [5]} & \cellcolor{gray!15}Minimal & \cellcolor{yellow!15}Low & \cellcolor{gray!15}Minimal & \cellcolor{gray!15}Minimal & \cellcolor{gray!15}Minimal & \cellcolor{green!25}High & \cellcolor{green!25}High \\
\hline
\textbf{Query Monitoring [6]} & \cellcolor{green!15}Medium & \cellcolor{green!25}High & \cellcolor{yellow!15}Low & \cellcolor{green!15}Medium & \cellcolor{green!15}Medium & \cellcolor{green!15}Medium & \cellcolor{green!15}Medium \\
\bottomrule
\end{tabular}

\begin{tablenotes}
    \scriptsize
    \item \textbf{Effectiveness Levels:} High (dark green) - Highly effective; Medium (light green) - Moderately effective; Low (yellow) - Limited effectiveness; Minimal (gray) - Minimal or no effectiveness.
    
    \begin{tabular}{@{}p{0.31\textwidth}p{0.31\textwidth}p{0.31\textwidth}@{}}
    \item \textbf{[1]} \citet{li2024translinkguard, li2024coreguard} & \item \textbf{[2]} \citet{wang2024guardemb}, \citet{pang2025modelshield} & \item \textbf{[3]} \citet{feng2024privacy}, \citet{patil2023can} \\
    \item \textbf{[4]} \citet{li2024llm}, \citet{wang2023self} & \item \textbf{[5]} \citet{he2022cater}, \citet{kim2024protection} & \item \textbf{[6]} \citet{wang2023self}
    \end{tabular}
\end{tablenotes}
\end{threeparttable}
\end{table*}


\subsection{Attack Effectiveness}

\noindent\textbf{Functional Similarity.} This metric assesses how closely an extracted model replicates the behavior of the target LLM, which focuses on capability transfer rather than exact parameter recovery. Functional similarity includes several metrics. Agreement rate measures the percentage of cases where extracted and target models produce equivalent outputs given identical inputs \cite{carlini2024stealing}. Behavioral consistency evaluates how reliably an extracted model reproduces specific patterns of the target model \cite{xu2021student}. Task-specific performance correlation quantifies alignment between extracted and target models on standardized benchmarks \cite{krishna2020thieves}. Perplexity similarity provides a continuous measure of functional extraction success by comparing cross-perplexity between models. This metric is particularly valuable when evaluating extraction against large generative models, which produce responses that make exact matching impractical.

\noindent\textbf{Data Recovery Rate.} This metric quantifies an attack's success in extracting sensitive or structured information, which is embedded within a target model. It encompasses several sub-metrics that evaluate different aspects of recovery success. Training data extraction success measures the percentage of training examples that are successfully recovered from a model's memorized content \cite{carlini2021extracting}. Precision and recall of extracted data evaluate both the accuracy and comprehensiveness of recovered information, which is particularly important when extracting structured data such as personally identifiable information (PII) \cite{huang2022are}. Private information exposure rate measures how effectively an attack extracts specifically targeted sensitive information, such as private user data or confidential records \cite{zhang2023ethicist}. Prompt recovery accuracy evaluates how accurately an attack reconstructs system prompts, which is assessed by measuring both semantic similarity and functional equivalence between the original and recovered prompts \cite{sha2024prompt}. These sub-metrics provide a comprehensive evaluation of an attack's ability to recover different types of information embedded within a model, highlighting the diverse risks posed by model extraction attacks.
\subsection{Defense Performance}

\noindent\textbf{Security Metrics.} These metrics evaluate how effectively defense mechanisms reduce the success of extraction attacks. They include attack prevention rate, which quantifies the reduction in extraction success when defensive measures are deployed \cite{li2024translinkguard}. Query detection accuracy measures a defense system's ability to identify malicious query patterns, which indicate extraction attempts \cite{wang2023self}. Extraction cost increase quantifies the additional computational and query resources that attackers must expend when defensive measures are in place \cite{pang2025modelshield}. Watermark robustness measures how well ownership signals persist in extracted models, which enables the detection of unauthorized clones \cite{wang2024guardemb}.

\noindent\textbf{Utility Metrics.} These metrics evaluate how defense mechanisms affect the model's ability to perform its intended tasks. They include performance preservation, which measures the degree to which defensive measures maintain original model capabilities \cite{li2024coreguard}. Response quality preservation quantifies changes in generation quality, which occur when defensive measures are applied \cite{he2022cater}. Computational overhead assesses the additional processing burden, which is introduced by defensive measures \cite{wang2023self}. False positive rate measures how frequently legitimate queries are incorrectly flagged or degraded, which occurs due to defensive measures \cite{kim2024protection}. Utility assessment enables practitioners to select defense strategies that provide adequate protection. At the same time, it ensures that the core value proposition of their deployed models is preserved.

\section{Limitations and Future Directions}

\noindent\textbf{Extraction Attack Perspectives.} Current attempts to replicate model functionality face challenges in deployment due to access restrictions and high computational requirements for proprietary models. Most attacks target isolated aspects of a model, which limits their comprehensiveness and adaptability to defenses. Future research should focus on combining multiple attack techniques to overcome specific defenses and optimizing query efficiency to reduce costs and avoid query restrictions.

\noindent\textbf{Defense Mechanism Advancements.} Existing defenses face challenges in deployment and effectiveness. Structural defenses require significant modifications, which limits their applicability to deployed systems, while output manipulation methods struggle to balance protection with generation quality. Most defenses lack formal security assurances and rely on empirical evaluation. Future defense research should prioritize defenses that can be applied to existing models without requiring structural changes or retraining. 

\section{Related Work}

Several surveys have explored various dimensions of security and privacy challenges for large language models. Comprehensive surveys like those by Das et al. \cite{das2025security} and Wang et al. \cite{wang2024unique} provide broad overviews of LLM security and privacy challenges, which include multiple types of threats and vulnerabilities. Yao et al. \cite{yao2024survey} organize their analysis into ``the good'' (beneficial security applications), ``the bad'' (offensive uses), and ``the ugly'' (inherent vulnerabilities), which provides a balanced perspective on LLMs' security implications. In contrast, other surveys focus on specific attack types or scenarios. Zhao et al. \cite{zhao2025survey} analyze attempts to replicate model functionality in distributed computing with some coverage of LLMs. Mathew \cite{mathew2024enhancing} specifically examines attacks that manipulate model behavior using malicious prompts and their defenses. Esmradi et al. \cite{esmradi2023comprehensive} survey implementation techniques and mitigation strategies for various attacks. Some surveys emphasize particular dimensions of LLM security. Liu et al. \cite{liu2025ethical} concentrate on societal and ethical consequences of LLM security, which focus more on societal impacts than technical mechanisms. Ma et al. \cite{ma2025safety} provide a safety-oriented perspective. Cui et al. \cite{cui2024recent} offer a more technically focused examination of recent attack and defense approaches. Our survey differs from these works by providing a systematic taxonomy of model extraction attacks specifically for LLMs. It examines the unique vulnerabilities of transformer-based architectures and offers a detailed analysis of extraction methods, such as recovering data used to train the model and prompt extraction, which exploit the distinctive characteristics of large language models.

\section{Conclusion}
This survey provides a structured analysis of attempts to replicate model functionality and the mechanisms to prevent them, which are critical for LLM security. We present a taxonomy, which categorizes attack types and defense mechanisms. Our framework includes evaluation metrics, which highlight the balance between security measures and model usability. We identify current limitations and research gaps, particularly in combining multiple attack techniques and establishing formal assurances of defense effectiveness. Finally, this work offers a structured foundation for developing robust protection strategies, which are essential for LLMs as critical commercial infrastructure.


\bibliographystyle{ACM-Reference-Format}
\bibliography{sample-base}










\end{document}